\documentclass[peer-reviewed]{aes133}
\usepackage{amsmath,bbold,xcolor}

\renewcommand{\t}{\textrm}
\renewcommand{\indent}{\hspace{20pt}}

\let\ifpdf\relax
 
\authors{Dennis L. Sun\aff{1,}\aff{2}, 
         Julius O. Smith III\aff{2}}

\affiliation[1]{Department of Statistics, Stanford University}
\affiliation[2]{Center for Computer Research in Music and Acoustics (CCRMA), Stanford University}

\correspondence{Dennis Sun}{dlsun@stanford.edu}

\lastnames{Sun, Smith}

\title{Estimating a Signal from a Magnitude Spectrogram via Convex Optimization}

\shorttitle{Convex Signal Estimation from a Magnitude Spectrogram}

\begin{abstract}
The problem of recovering a signal from the magnitude of its short-time Fourier transform (STFT) is a longstanding one in audio signal processing. Existing approaches rely on heuristics which often perform poorly because of the nonconvexity of the problem. We introduce a formulation of the problem that lends itself to a tractable convex program. We observe that our method yields better reconstructions than the standard Griffin-Lim algorithm. We provide an algorithm and practical implementation details, including a discussion of how the method can be scaled up to larger examples.
\end{abstract}

\begin{document}

\maketitle % MANDATORY! 

\section{Introduction}

The problem of estimating a signal from the magnitude of its short-time Fourier transform (STFT) has vexed audio signal processing researchers for decades \cite{HayesLimOpp80,NawabQuatieriLim83,Slaney94}. The fundamental challenge can be summarized as:

\begin{quote}
\emph{Fourier coefficients are complex numbers, but in many applications we only know their magnitudes and not their phases.}
\end{quote}

For example, in many frequency-domain audio processing techniques such as time-scale modification and speech enhancement, it is typically straightforward to modify the magnitudes of the Fourier coefficients but not their phases. In the case of time stretching, it is clear that the energy (i.e., magnitudes) should be spread over more time bins, but unclear how to specify the phases accordingly.

The absence of phase information is problematic when we want to reconstruct the modified signal, for which it is necessary to invert the STFT---an operation which requires the full complex coefficients (i.e., both magnitudes and phases). Hence, the magnitude-only reconstruction problem is also termed \emph{phase retrieval} in the literature. Using incompatible phase information in the reconstruction can result in amplitude-modulation distortion \cite{LarocheDolson97}.

Recently, interest in the magnitude-only reconstruction problem has resurged in the source separation community, where matrix factorization has become a standard workhorse \cite{SmaragdisBrown03}. Since these approaches operate entirely on the magnitude spectrogram, the same difficulties arise in reconstructing the time-domain signals. A heuristic that is commonly used in this literature is to set the phases of each of the recovered signals equal to the phases of the original mixture signal \cite{Smaragdis10}. Some justification has been offered for this approach; under certain modeling assumptions, it is a consequence of the Wiener filter reconstruction \cite{Fevotte09}.

Griffin and Lim \cite{GL84} proposed a more general solution to the magnitude-only reconstruction problem. As their approach is now the standard, we briefly review their algorithm in the next section so that we may later contrast it with our approach.

\section{Existing Approaches}

Let $|Y_w(mR,\omega)|$ denote the desired magnitude STFT, where $R$ is the hop size, $w$ denotes the window, and $m$ and $\omega$  index the time frame and frequency, respectively. Our goal is to find a signal $x$, whose magnitude STFT $|X_w(mR,\omega)|$ is as close to the desired as possible, in a least squares sense:
\begin{equation}
\sum_{m=-\infty}^\infty \int_{\omega=-\pi}^\pi \left( |X_w(mR,\omega)| - |Y_w(mR,\omega)| \right)^2\,d\omega
\label{obj}
\end{equation}

The Griffin-Lim algorithm attempts to minimize this objective by starting with an initial guess for the signal $x^{(0)}$ and iterating the following steps until convergence:

\begin{enumerate}

\item Compute the STFT of the current signal estimate $x^{(i)}$:
\begin{equation}
X_w^{(i)}(mR,\omega) := \sum_{n=-\infty}^\infty w(mR-n) x^{(i)}(n) e^{-j\omega n}.
\end{equation}
Retain the phases of $X_w^{(i)}$, but replace its magnitudes by the desired magnitudes:
\begin{equation}
X_w^{(i+1)}(mR,\omega) := |Y_w(mR,\omega)|\frac{X_w^{(i)}(mR,\omega)}{|X_w^{(i)}(mR,\omega)|}.
\end{equation}

\item Compute the IDTFT of $X_w^{(i+1)}(mR,\cdot)$, denoted $\hat x_{w, mR}^{(i+1)}$, for each $m$.\footnote{Ideally we would like for $\hat x_{w, mR}^{(i+1)}$ to be the windowed version of some signal $x$, i.e. 
\begin{equation}
\hat x_{w, mR}^{(i+1)}(n) = w(mR-n)x(n).
\label{gl2a}
\end{equation} 
To see that this is not always possible, consider the case where $\hat x_{w,mR}(n) \neq 0$ but $w(mR-n) = 0$. Furthermore, we would need a single signal $x$ such that (\ref{gl2a}) holds for all $m$.} Then, find a signal $x$ such that the windowed version of $x$, $x_{w,mR}$, is close to $\hat x_{w,mR}$ for each $m$ in a least-squares sense. The solution turns out to be:
\begin{equation}
x^{(i+1)} := \frac{\sum_{m=-\infty}^\infty w(mR-n)\hat x_{w,mR}^{(i+1)}(n)}{\sum_{m=-\infty}^\infty w^2(mR-n)}.
\end{equation}

\end{enumerate}

Griffin and Lim showed that this algorithm decreases the objective (\ref{obj}) on each iteration and converges to a stationary point. However, the objective is nonconvex, so different initializations $x^{(0)}$ can lead to different solutions, and there are no guarantees on convergence to a global optimum. In fact, as we will show, the algorithm often fails to recover a signal from its true magnitude spectrogram, even when several initializations are used.

It is worth noting that the magnitude-only reconstruction problem has been formulated in other ways: as a probabilistic model \cite{AchanRoweis03} and as a root-finding problem \cite{BouvrieEzzat06}. However, these methods suffer from the same drawbacks as the Griffin-Lim algorithm---namely, that they try to solve a nonconvex problem and hence are prone to local optima. To our knowledge, the algorithm that we describe in the next section is the first convex formulation for this problem.

\section{Convex Formulation}

The Griffin-Lim algorithm resembles a number of earlier alternating projections algorithms \cite{GS72} for solving the basic phase retrieval problem (i.e., where the magnitudes are of only a single Fourier transform). Recently it was shown in \cite{Candes11a} that this basic problem could be recast in terms of estimating a rank-1 matrix $X$ which is the outer product of the signal with itself, i.e., $X = xx^*$. Once $X$ is obtained, the putative signal can be recovered by factorizing $X$, provided that it is rank one. 

This technique of \emph{lifting} the problem of estimating a vector $x$ to estimating a matrix $X = xx^*$ is standard in the optimization literature \cite{AGJL05}. Although it drastically increases the dimensions of the problem, it leads naturally to a tractable convex program, as we shall see. In this section, we formulate this problem in the context of the STFT. In the spirit of \cite{Candes11a}, who term their algorithm PhaseLift, we call ours STliFT.

First, we consider the problem of recovering a signal from its true magnitude spectrogram (i.e., when a signal with that exact magnitude spectrogram exists). Suppose we use an FFT of length $N$ for each frame. Letting $s_k(n) = e^{j2\pi n k/N}$, $k=0,...,N-1$ denote the sampled complex sinusoids corresponding to this FFT length, the $k^\t{th}$ Fourier coefficient for the $m^\t{th}$ time frame is $\langle x_{w,mR}, s_k\rangle$, where $x_{w,mR}$ denotes the windowed signal $x_{w,mR}(n) = w(mR-n)x(n)$, zero-padded to the FFT length $N$. Since we want the magnitude of each of these coefficients to be $|Y_w(mR,\omega_k)|$, the problem can be recast as:
\begin{equation}
\begin{aligned}
& \text{find} & & x \\
& \text{subject to} & & \left|\langle x_{w,mR}, s_k \rangle\right| = |Y_w(mR,\omega_k)| \\
& & & \indent 0 \leq k < N,\,\, 0 \leq m < T
\end{aligned}
\end{equation}
Next, we note that we can write the windowed signal $x_{w,mR}$ as a matrix product $W_{mR} x$, where $W_{mR}$ is of the form:
\[ W_{mR} = \begin{pmatrix}  
& & & w(-\frac{N}{2}+1) & & & \\
 & 0 & & & \ddots & & & 0 &\\
& & & & & w(\frac{N}{2}) & 
\end{pmatrix}.
 \]
Now we square both sides of the equality constraint. Using the fact that $x_{w,mR} = W_{mR}x$, we can write the left-hand side as
\begin{align*}
\left|\langle x_{w,mR}, s_k \rangle\right|^2 &=  \t{tr}((W_{mR} x)^T s_k s_k^* (W_{mR} x)) \\
&= \t{tr}( \underbrace{(W_{mR}^T s_k) (W_{mR}^T s_k)^*}_{S_{k,m}} \underbrace{x x^T}_X)
%&= \t{tr}( S_{k,m} X )
\end{align*}
where in the second step we have used the trace identity $\t{tr}(AB) = \t{tr}(BA)$ to group $x$ and $x^T$ into a single rank-1 matrix $X$. Hence the problem becomes:
\begin{equation}
\begin{aligned}
& \text{find} & & X \\
& \text{subject to} & & \t{rank}(X) = 1,\, X \succeq 0 \\
& & & \t{tr}( S_{k,m} X ) = |Y_w(mR,\omega_k)|^2 \\
& & & \indent 0 \leq k < N,\,\, 0 \leq m < T
\end{aligned}
\label{rank}
\end{equation}
This is equivalent to the original problem and thus also nonconvex, except now the nonconvexity plainly arises from a rank constraint. We can relax (\ref{rank}) to a convex problem by replacing the rank constraint with its convex surrogate, a trace minimization:
\begin{equation}
\begin{aligned}
& \text{minimize} & & \t{tr}(X) \\
& \text{subject to} & & X \succeq 0 \\
& & & \t{tr}( S_{k,m} X ) = |Y_w(mR,\omega_k)|^2 \\
& & & \indent 0 \leq k < N,\,\, 0 \leq m < T
\end{aligned}
\label{noiseless}
\end{equation}
This problem is not only convex; it is in fact the dual semidefinite program (SDP) described in \cite{BL96}.

Now we consider the case where there is not necessarily any signal with the specified magnitude spectrogram, and we wish to find a signal whose magnitude spectrogram comes as close to the specification as possible. Because the above problem is formulated for the \emph{power} spectrogram rather than the \emph{magnitude} spectrogram, it cannot readily be adapted to minimize the distance between the magnitude spectrograms as in (\ref{obj}). However, it is straightforward to extend the above framework to minimize the distance between the power spectrograms, by solving
\begin{equation}
\begin{aligned}
& \text{minimize} & & \sum_{k,m} (\t{tr}(S_{k,m}X)-|Y_w(mR,\omega_k)|^2)^2 \\
& & & \indent+ \lambda\cdot\t{tr}(X) \\
& \text{subject to} & & X \succeq 0 \\
\end{aligned}
\label{noisy}
\end{equation}
where $\lambda$ controls the tradeoff between minimizing the distance between the spectrograms and the low-rank constraint on $X$. In practice, we initialize $\lambda$ large enough so that the solution is $X = 0$ and solve (\ref{noisy}) repeatedly for decreasing values of $\lambda$ until we obtain an (approximately) rank-1 solution.

Before concluding this discussion of the model, we comment that Euclidean distance may be inappropriate as a measure of divergence between the estimated and desired spectrogram. Euclidean distance is symmetric on a linear scale, so if the desired power in a time-frequency bin is $0.1$, then Euclidean distance would favor a reconstruction which assigns a power of $0.01$ to that bin, over one which assigns $0.2$. However, the human ear perceives energy on a logarithmic scale; $0.01$ represents a discrepancy of $20$ dB, whereas $0.2$ represents one of only $6$ dB. For these reasons, logarithmic divergences such as Kullback-Leibler or Itakura-Saito may be preferable to Euclidean distance for audio \cite{Fevotte09}. 

For purposes of comparison with Griffin-Lim, we only deal with the Euclidean case in this work. However, our formulation above can be readily extended to these other divergences by simply replacing $(\t{tr}(S_{k,m}X)-|Y_w(mR,\omega_k)|^2)^2$ by the appropriate divergence $D(\t{tr}(S_{k,m}X),|Y_w(mR,\omega_k)|^2)$. Such problems can be solved using the algorithm described in the next section, by modifying the gradient $\nabla L$ appropriately.

\section{Practical Implementation}

In this section, we describe a practical algorithm for solving (\ref{noisy}). Although the problem can in principle be solved using standard solvers, the scale of the problem ($X$ is an $n\times n$ matrix, where $n$ is the length of the signal) makes this all but impossible for any real-world problem.

A projected gradient descent algorithm was proposed to solve the PhaseLift problem \cite{Candes11a}. Letting $L(X)$ denote the objective function in (\ref{noisy}), the algorithm starts with an initial guess $X_0$ and iterates:
\[ X_{k+1} = \mathcal{P}(X_k - t_k\nabla L(X_k)) \]
where $\mathcal{P}$ denotes projection onto the feasible set $X \succeq 0$, $\nabla L$ the gradient, and $t_k$ a ``suitable'' step size. $\mathcal{P}$ performs an eigendecomposition on the matrix $X$ and thresholds all negative eigenvalues to 0. 

For Euclidean distance as in (\ref{noisy}), the gradient is:
\[ \nabla L(X) =  2\sum_{k,m} (\t{tr}(S_{k,m}X) - |Y_w(mR,\omega_k)|^2)S_{k,m} + \lambda I \]
We emphasize that this is a conceptual, rather than a practical, formula. In practice, the structure of each $S_{k,m}$ (its nonzero entries form a block within the $n\times n$ matrix) should be exploited for fast computation. Such optimizations are implemented in the Matlab code for this algorithm, which will be made available at the first author's webpage.\footnote{{\tt http://www-stat.stanford.edu/$\sim$dlsun}}

Although gradient descent scales well to large problems, it can converge slowly for ill-conditioned problems, where the direction of steepest descent can be nearly orthogonal to the direction of the minimum. This results in the characteristic ``zig-zag'' trajectory of gradient descent \cite{BV}. One way to accelerate the convergence is to evaluate the gradient not at $X_k$ but at an auxiliary point $Y_k$ which is a function of the past values of $X$, in order to capture ``momentum'' of the trajectory. \cite{Candes11a} suggests a variant of the FISTA acceleration scheme \cite{FISTA}:
\begin{align}
X_{k+1} &= \mathcal{P}(Y_k - t_k \nabla L(Y_k)) \nonumber\\
\theta_{k+1} &= 2 \left( 1 + \sqrt{1 + 4/\theta_k^2} \right)^{-1} \nonumber \\
\beta_{k+1} &= \theta_{k+1}(\theta^{-1}_k - 1) \nonumber\\
Y_{k+1} &= X_{k+1} + \beta_k (X_{k+1} - X_k)
\label{nesterov}
\end{align}
We have found this algorithm to scale to reasonably large problems and have confirmed that the accelerations defined by (\ref{nesterov}) drastically improve the performance of projected gradient descent in our setting. 

The final piece of the algorithm is the step size $t_k$. Although $t_k$ can be chosen using a line search, evaluating the objective is prohibitively expensive for large problems since it involves an eigendecomposition. We have found choosing a fixed step size by trial and error to work well in practice. For step sizes that are too large, the objective values diverge within a few iterations, so we use the largest step size for which the objective values still converge.

\section{Results}

First, we consider the ``noiseless'' problem of recovering a signal from its true magnitude STFT. We first generated 20 random signals each of length $n=16$ and $n= 32$. We then determined the magnitude STFT of each signal using Hann windows of length $M=5,7,9,11$ for both the $n=16$ and $n=32$ signals, and also windows of length $M=13,15,17,19,21$ for the $n=32$ signals. We considered all hop sizes that resulted in constant overlap-add for each window length. The FFT size was taken to be the same as the signal length: $N=n$.

We then applied the Griffin-Lim algorithm and the program (\ref{noiseless}) to each magnitude spectrogram to obtain a signal estimate. For Griffin-Lim, we used 10 random initializations and retained only the ``best'' (see below) of the 10 initializations. To evaluate performance, we tabulated the percentage of the 20 signals whose estimates achieved a value of (\ref{obj}) less than $10^{-3}$. Results are presented in Table \ref{comp16}.

\begin{table}

{\small\begin{tabular}{cccc}
\hline
\multicolumn{4}{c}{$n=16$} \\
\hline
\parbox{.67in}{Window Length ($M$)} & \parbox{.47in}{Hop Size ($R$)} & \parbox{.6in}{Accuracy (\%) of G-L} & \parbox{.6in}{Accuracy (\%) of (\ref{noiseless})} \\
\hline
5 & 2 & 0 & 65 \\
5 & 1 & 30 & 100 \\
\hline
7 & 3 & 0 & 50 \\
7 & 2 & 10 & 100 \\
7 & 1 & 60 & 100 \\
\hline
9 & 4 & 5 & 60 \\
9 & 2 & 70 & 100 \\
9 & 1 & 80 & 100 \\
\hline
11 & 5& 5 & 60 \\
11 & 2 & 95 & 100 \\
11 & 1 & 95 & 100 \\
\hline
\hline
\multicolumn{4}{c}{$n=32$} \\
\hline
\parbox{.67in}{Window Length ($M$)} & \parbox{.47in}{Hop Size ($R$)} & \parbox{.6in}{Accuracy (\%) of G-L} & \parbox{.6in}{Accuracy (\%) of (\ref{noiseless})} \\
\hline
5 & 2 & 0 & 20 \\
5 & 1 & 0 & 100 \\
\hline
7 & 3 & 0 & 15 \\
7 & 2 & 0 & 100 \\
7 & 1 & 15 & 100 \\
\hline
9 & 4 & 0 & 15 \\
9 & 2 & 10 & 100 \\
9 & 1 & 20 & 100 \\
\hline
11 & 5& 0 & 20 \\
11 & 2 & 55 & 100 \\
11 & 1 & 55 & 100 \\
\hline
13 & 6 & 0 & 55 \\
13 & 4 & 0 & 100 \\
13 & 3 & 45 & 100 \\
13 & 2 & 80 & 100 \\
13 & 1 & 65 & 100 \\
\hline
15 & 7 & 0 & 55 \\
15 & 2 & 85 & 100 \\
15 & 1 & 80 & 100 \\
\hline
17 & 8 & 0 & 60 \\
17 & 4 & 60 & 100 \\
17 & 2 & 80 & 100 \\
17 & 1 & 75 & 100 \\
\hline
19 & 9 & 0 & 60 \\
19 & 6 & 0 & 100 \\
19 & 3 & 85 & 100 \\
19 & 2 & 85 & 100 \\
19 & 1 & 90 & 100 \\
\hline
21 & 10 & 0 & 20 \\
21 & 5 & 65 & 100\\
21 & 4 & 90 & 100 \\
21 & 2 & 85 & 100 \\
21 & 1 & 100 & 100 \\
\hline
\end{tabular}}

\caption{\small Comparison of Griffin-Lim and program (\ref{noiseless}) for different configurations in the ``noiseless'' setup. Accuracy is the percentage of the 20 test signals for which the objective (\ref{obj}) of the solution was $< 10^{-3}$.}
\label{comp16}
\end{table}

\begin{figure}

\includegraphics[width=.5\textwidth]{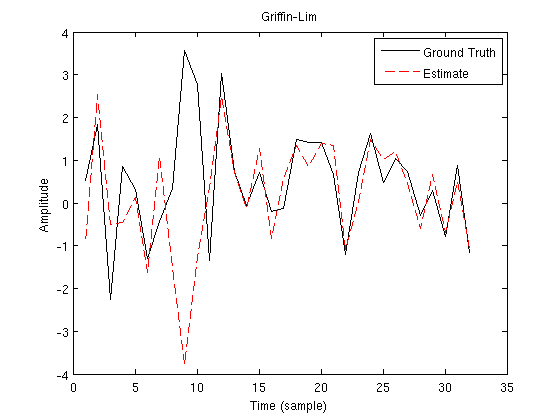}

\includegraphics[width=.5\textwidth]{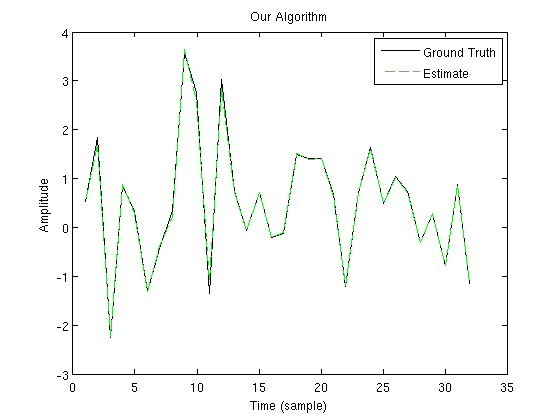}

\caption{We applied the Griffin-Lim algorithm and program (\ref{noiseless}) to a random signal of length $n=32$ with a window of length $M=15$ and a hop size of $R=7$. Shown at top is an instance where Griffin-Lim fails to converge to the global optimum of (\ref{obj}) and returns an unreasonable solution. Shown at bottom: our algorithm basically recovers the signal from the magnitude spectrogram, up to the numerical precision of the algorithm.}

\label{signals}

\end{figure}

We found that our algorithm recovers the signal more consistently than the Griffin-Lim algorithm. Furthermore, the objective values attained by the latter when it did not converge to the optimum were much larger than $1$, which empirically supports the claim that alternating projections algorithms have a tendency to get stuck in local optima for nonconvex problems. Recovery is a problem for both algorithms when the hop size is large, but especially for Griffin-Lim, most likely because there are fewer constraints with a large hop size, so the landscape is dotted with more local optima.

Next, we consider a more realistic ``noisy'' setup in which there may not be any signal with the given magnitude STFT and we wish only to find a signal minimizing (\ref{obj}). We take the same 20 signals as before, calculate their power spectrogram, and add i.i.d. Gaussian noise with mean 0 and standard deviation 0.2. (We set to zero any entries which end up negative as a result.) We then apply the Griffin-Lim algorithm and the program (\ref{noisy}) to this modified magnitude STFT, and compare the values of (\ref{obj}) for the two estimates. Results are presented in Table \ref{comp32}.

Here the advantages of our algorithm are not so clear cut. Griffin-Lim tends to perform better for large hop sizes, whereas our algorithm does better for smaller hop sizes. However, the two algorithms appear to become increasingly similar as window length increases.

\begin{table}

{\small\begin{tabular}{cccc}
\hline
\multicolumn{4}{c}{$n=16$} \\
\hline
\parbox{.67in}{Window Length ($M$)} & \parbox{.47in}{Hop Size ($R$)} & \parbox{.74in}{Median Rel.\\\% Err.\,of G-L} & \parbox{.7in}{Median Rel.\\\% Err.\,of (\ref{noisy})} \\
\hline
5 & 2 & 1.18 & 5.66 \\
5 & 1 & 0.82 & 0.86\\
\hline
7 & 3 & 1.25 & 4.27 \\
7 & 2 & 1.85 & 1.00 \\
7 & 1 & 0.44 & 0.35 \\
\hline
9 & 4 & 1.03 & 4.32 \\
9 & 2 & 0.66 & 0.41\\
9 & 1 & 0.31 & 0.27 \\
\hline
11 & 5& 0.85 & 5.60 \\
11 & 2 & 0.35 & 0.31 \\
11 & 1 & 0.20 & 0.21\\
\hline
\hline
\multicolumn{4}{c}{$n=32$} \\
\hline
\parbox{.67in}{Window Length ($M$)} & \parbox{.47in}{Hop Size ($R$)} & \parbox{.74in}{Median Rel.\\\% Err.\,of G-L} & \parbox{.7in}{Median Rel.\\\% Err.\,of (\ref{noisy})} \\
\hline
5 & 2 & 0.89 & 16.7 \\
5 & 1 & 1.72 & 0.22 \\
\hline
7 & 3 & 1.00 & 3.91 \\
7 & 2 & 2.10 & 0.24 \\
7 & 1 & 0.77 & 0.12 \\
\hline
9 & 4 & 0.46 & 7.80 \\
9 & 2 & 1.43 & 0.11 \\
9 & 1 & 1.07 & 0.06 \\
\hline
11 & 5& 0.52 & 8.65 \\
11 & 2 & 0.74 & 0.09 \\
11 & 1 & 0.67 & 0.05 \\
\hline
13 & 6 & 0.40 & 9.82 \\
13 & 4 & 1.06 & 0.15 \\
13 & 3 & 1.15 & 0.12 \\
13 & 2 & 0.39 & 0.08 \\
13 & 1 & 0.05 & 0.04 \\
\hline
15 & 7 & 0.40 & 4.62 \\
15 & 2 & 0.07 & 0.07 \\
15 & 1 & 0.04 & 0.04 \\
\hline
17 & 8 & 0.39 & 5.77 \\
17 & 4 & 0.12 & 0.11 \\
17 & 2 & 0.06 & 0.06 \\
17 & 1 & 0.04 & 0.04 \\
\hline
19 & 9 & 0.31 & 3.08 \\
19 & 6 & 0.93 & 0.14 \\
19 & 3 & 0.09 & 0.08 \\
19 & 2 & 0.05 & 0.05 \\
19 & 1 & 0.04 & 0.03 \\
\hline
21 & 10 & 0.51 & 4.83 \\
21 & 5 & 0.15 & 0.14\\
21 & 4 & 0.09 & 0.09 \\
21 & 2 & 0.05 & 0.05 \\
21 & 1 & 0.03 & 0.03 \\
\hline
\end{tabular}}

\caption{\small Comparison of Griffin-Lim and program (\ref{noisy}) for different configurations in the ``noisy'' setup. The median relative error (defined as (\ref{obj}) divided by the total power $\sum_{m,k} |Y_w(mR,\omega_k)|^2$) over 10 test signals is shown.}
\label{comp32}

\end{table}

\section{Discussion}

Although the algorithm is promising on small examples, it remains to scale it to real audio data. To appreciate the difficulties of scaling the algorithm, note that the algorithm estimates an $n \times n$ matrix, where $n$ is the length of the audio clip. For a 3-second clip sampled at 44.1 kHz, this amounts to estimating a $10^5 \times 10^5$ matrix. 

We make several observations which may be useful for scaling the algorithm. First, we note that the algorithm is entirely parallelizable in the sense that we can divide the signal into smaller segments and run the algorithm separately on each chunk. The only thing that we lose is the information that would have been provided by the time frames that stretch across two segments. Furthermore, it may not be necessary to compute the phase over the entire signal. Since phase is not audible over stationary stretches, one could apply the algorithm only on the transient time frames and allow phase to ``free run'' elsewhere. Since transients are short in duration and often involve smaller windows, these problems will be much smaller in scale.

\section{Conclusion}

This paper has introduced a novel method for estimating a signal from its magnitude STFT. Although it remains to scale the algorithm to real-life audio applications, our approach is highly competitive with existing algorithms in stylized simulations. 

\section{Acknowledgements}

We would like to thank Rahul Mazumder for helpful discussions.


\begin{thebibliography}{99}

\bibitem{HayesLimOpp80}
M. H. Hayes, J. S. Lim, and A. V. Oppenheim, ``Signal Reconstruction from Phase or Magnitude,''  \emph{IEEE Trans. on Acoust., Speech, Signal Processing} \textbf{28}, pp. 672-680, August 1980.

\bibitem{NawabQuatieriLim83}
S. H. Nawab, T. F. Quatieri, and J. S. Lim, ``Signal Reconstruction from Short-Time Fourier Transform Magnitude,'' \emph{IEEE Trans. on Acoust., Speech, Signal Processing} \textbf{31}, pp. 986-998, August 1983.

\bibitem{Slaney94}
M. Slaney, D. Naar, and R. F. Lyon, ``Auditory Model Inversion for Sound Separation,'' \emph{Proc. IEEE Int. Conf. on Acoust., Speech, and Signal Processing (ICASSP)}, April 1994.

\bibitem{LarocheDolson97}
J. Laroche and M. Dolson, ``About this phasiness business,'' \emph{Proc. IEEE Workshop on Applications of Signal Processing to Audio and Acoustics (WASPAA)}, pp. 91-94, October 1997.

\bibitem{SmaragdisBrown03}
P. Smaragdis and J. C. Brown, ``Non-negative matrix factorization for polyphonic music transcription,'' \emph{Proc. IEEE Workshop on Applications of Signal Processing to Audio and Acoustics (WASPAA)}, October 2003.

\bibitem{Smaragdis10}
P. Smaragdis, B. Raj, and M. Shashanka, ``Missing data imputation for time-frequency representations of audio signals,'' \emph{J. Signal Processing Systems}, August 2010.

\bibitem{Fevotte09}
C. F\'{e}votte, N. Bertin, J.-L. Durrieu, ``Nonnegative Matrix Factorization with the Itakura-Saito Divergence: With Application to Music Analysis,'' \emph{Neural Computation} \textbf{21}, pp. 793-830, 2009.

\bibitem{GL84}
D. W. Griffin and J. S. Lim, ``Signal Estimation from Modified Short-Time Fourier Transform,'' \emph{IEEE Trans. on Acoust., Speech, Signal Processing} \textbf{32}, pp. 236-243, April 1984.

\bibitem{AchanRoweis03}
K. Achan, S. T. Roweis, and B. J. Frey, ``Probabilistic Inference of Speech Signals from Phaseless Spectrograms,'' \emph{Advances in Neural Information Processing Systems (NIPS)} \textbf{16}, MIT Press, 2004.

\bibitem{BouvrieEzzat06}
J. Bouvrie and T. Ezzat, ``An Incremental Algorithm for Signal Reconstruction from Short-Time Fourier Transform Magnitude,'' \emph{Proc. Ninth International Conference on Spoken Language Processing (Interspeech)}, September 2006.

\bibitem{GS72}
R.W. Gerchberg and W.O. Saxton, ``A practical algorithm for the determination of phase from image and diffraction plane pictures,'' \emph{Optik} \textbf{35}, pp. 237-246, 1972.

\bibitem{Candes11a}
E. J. Cand\`{e}s, Y. Eldar, T. Strohmer and V. Voroninski, ``Phase retrieval via matrix completion,'' Preprint.

\bibitem{AGJL05}
A. D'Aspremont, L. El Ghaoui, M. I. Jordan, and G. R. G. Lanckriet, ``A Direct Formulation for Sparse PCA Using Semidefinite Programming,'' \emph{SIAM Review} \textbf{49}, pp. 434-448, 2005.

\bibitem{BL96}
S. Boyd and L. Vandenberghe, ``Semidefinite Programming,'' \emph{SIAM Review} 38, pp. 49-95, March 1996.

%\bibitem{CVX}
%M. Grant and S. Boyd, CVX: Matlab software for disciplined convex programming, version 1.21, {\tt http://cvxr.com/cvx}, April 2011.
%
%\bibitem{SASP}
%J. O. Smith, \emph{Spectral Audio Signal Processing}, W3K Publishing, 2011.

\bibitem{BV}
S. Boyd and L. Vandenberghe, \emph{Convex Optimizaton}, Cambridge University Press, 2004.

\bibitem{FISTA}
A. Beck and M. Teboulle. ``A Fast Iterative Shrinkage-Thresholding Algorithm for Linear Inverse Problems,'' \emph{Siam J. Imaging Sciences} \textbf{2}, pp. 183-202.

%\bibitem{Candes11b}
%E. J. Cand\`{e}s, T. Strohmer and V. Voroninski, ``PhaseLift: exact and stable signal recovery from magnitude measurements via convex programming,'' To appear in \emph{Comm. on Pure and Applied Mathematics}.



%\bibitem{DEK2}
%D. E. Knuth, {\it Selected papers on analysis of algorithms}, CSLI
%Publ., Stanford, CA, 2000; CNO
%CMP 1 762 319 
%
%\bibitem{DEK3}
%D. E. Knuth, Algorithmica {\bf 22} (1998), no.~4, 561--568; MR
%2000j:68037 
%
%\bibitem{DEK4}
%R. L. Graham, D. E. Knuth and O. Patashnik, {\it Concrete mathematics}
%(Polish), Translated from the
%second English (1994) edition by P. Chrzastowski, A. Czumaj,
%L. Gasieniec and M. Raczunas, Second
%edition, Wydawnictwo Naukowe PWN, Warsaw, 1998; MR 99m:68002
%
\end{thebibliography}
\end{document}